\documentclass[12pt]{article}
\usepackage{ifpdf}
\usepackage{graphics}
\usepackage{graphicx}
\usepackage{amsmath,mathptmx,amssymb,bm}
\newtheorem{theorem}{Theorem}

\DeclareGraphicsExtensions{.png,PNG,.pdf,.PDF}

\begin{document}


\title{Lie symmetries, conservation laws and exact solutions of a generalized quasilinear KdV equation with degenerate dispersion}


\author{M.S. Bruz\'on${}^{a}$, E. Recio${}^{b}$, T.M. Garrido${}^{c}$, R. de la Rosa${}^{d}$\\
 ${}^{a}$ Universidad de C\'adiz, Spain  (e-mail: m.bruzon@uca.es ). \\
 ${}^{b}$ Universidad de C\'adiz, Spain (e-mail:  elena.recio@uca.es). \\
 ${}^{c}$ Universidad de C\'adiz, Spain (e-mail:  tamara.garrido@uca.es). \\
 ${}^{d}$ Universidad de C\'adiz, Spain  (e-mail: rafael.delarosa@uca.es ). \\  
}

%
%

\date{}
 
\maketitle

\begin{abstract}

We provide a complete classification of point symmetries and low-order local conservation laws of the generalized quasilinear KdV equation in terms of the arbitrary function. The corresponding interpretation of symmetry transformation groups are given. In addition, a physical description of the conserved quantities is included. Finally, few travelling wave solutions have been obtained.\\

\end{abstract}



\maketitle

\section{Introduction}
Non-linear dispersive partial differential equations (PDEs) model a wide range of physical behaviour and phenomena. It is important to differentiate the dispersive effects of these equations. There are linear dispersive effects, non-linear dispersive effects and equations whose dispersive effects could vanish.

The well-known Korteweg de Vries (KdV) equation
$$u_t+u u_x + u_{xxx}=0, $$
is a non-linear dispersive PDE whose dispersive term, $u_{xxx}$, is linear. Hence, this term establishes a linear dispersion in the dynamics of this equation.

In the same way, there are dispersive equations in which the term which generates the dispersive effects is itself non-linear.
With regard to the third case, there are problems in which the dispersive effects may vanish. Such equations are called degenerate dispersive equations \cite{Amb}.

The following quasilinear KdV equation
\begin{equation}\label{edp}
u_t+ (u(uu_x)_x+\mu u^3)_x=0,
\end{equation}
is an example with degenerate dispersion that appeared in \cite{coop,ros}.

In this paper, we have studied a generalization of equation (\ref{edp}) given by
\begin{equation}\label{genedp}
u_t+ (u(uu_x)_x+f(u))_x=0,
\end{equation}
with $f'(u)\neq0$.

The research of Lie symmetries of PDEs may allow to construct invariant solutions, reduce the number of independent variables, reduce the PDE to ordinary differential equations (ODE) or reduce the order of the ODE, among others. Therefore, symmetries have become a powerful tool in the study of PDE \cite{bru1,bru2,bru3,recio}.

Lie method is a systematic method that allows to obtain all the Lie symmetries for a given equation, including those that only occur for some specific function $f$. When we are searching for symmetries the method will provide a set of special forms for the unknown function $f$ for which the equation admits symmetries.

Conservation laws are continuity equations which yield conserved physical quantities and conserved norms for all solutions of a given evolution equation \cite{blumanchevianco,PJO}. Their applications include checking the integrability of a PDE and verifying the accuracy of numerical methods for solutions.

There is a systematic method for finding all local conservation laws admitted by any given evolution equation, namely, the multiplier method \cite{Alo,PJO}. We apply this method to the generalized quasilinear KdV equation \eqref{genedp} by setting up and solving an overdetermined linear system of determining equations, which Anco and Bluman showed \cite{Alo,anco0,anco1,anco,ancorev} it is an adjoint version of the standard method for finding all symmetries of any given evolution equation.

Moreover, the study of Lie symmetries and conservation laws of equation (\ref{edp}) is also motivated by the research in \cite{ger}. It is presented that equation (\ref{edp}) admits a Hamiltonian structure, it has translation, reflection and scaling symmetries, and it conserves besides the Hamiltonian, the mass and the momentum.

Consequently, for the generalized equation \eqref{genedp} it is interesting to focus on the study of Lie symmetries and conservation laws in order to determine special cases for the arbitrary function, $f(u)$, with extra symmetries or conservation laws.

Finally, we seek for special cases of the arbitrary function that allow us to obtain exact solutions with physical interest.

\section{Point Symmetries}
\label{S1}

Let consider the one-parameter Lie group of infinitesimal transformations on the space of independent and dependent variables
\begin{equation}\label{group}
\begin{array}{lcl}
\hat{t}(t,x,u;\epsilon)&=&t+\epsilon \, \tau(t,x,u)+O(\epsilon ^2), \\
\hat{x}(t,x,u;\epsilon)&=&x+\epsilon \, \xi(t,x,u)+O(\epsilon^2), \\
\hat{u}(t,x,u;\epsilon)&=&u+\epsilon\, \eta(t,x,u)+O(\epsilon ^2),
\end{array}
\end{equation}

\noindent where $\epsilon$ represents the group parameter. Each infinitesimal point symmetry (\ref{group}) constitutes a generator
\begin{equation}\label{vect1}X=
\tau(t,x,u)\partial_t+ \xi(t,x,u) \partial_x+\eta(t,x,u)\partial_u.\end{equation}

We recall (\ref{vect1}) is a point symmetry of equation (\ref{genedp}) if the third order prolongation of (\ref{vect1}) leaves invariant equation (\ref{genedp}). In order to avoid unnecessary calculations, prolongations can be taken into account from a geometric point of view \cite{blumanchevianco,PJO}. In this way, the action of (\ref{vect1}) on the solution space of equation (\ref{genedp}) is similar to the action of the symmetry in characteristic form
$$
\hat{X}= \hat{\eta}\partial_u, \quad \hat{\eta}=\eta-\tau u_t-\xi u_x,
$$

\noindent Thus, by applying the symmetry invariance condition
\begin{equation}\label{newprolong}
pr^{(3)}\hat{X}(u_t+ (u(uu_x)_x+f(u))_x)=0 \quad \mbox{when} \quad u_t+ (u(uu_x)_x+f(u))_x=0,
\end{equation}
point symmetries can be determined. Here,
$$pr^{(3)}\hat{X}=\hat{X}+ \left(D_x \hat{\eta} \right)\partial_{u_{x}}+ \left(D_t \hat{\eta} \right)\partial_{u_{t}}+ \left(D_x^2 \hat{\eta} \right)\partial_{u_{xx}}+ \left(D_x^3 \hat{\eta} \right)\partial_{u_{xxx}},$$
represents the third prolongation of the generator $\hat{X}$ whereas $D_t$ and $D_x$ represent the total derivatives respect to $t$ and $x$ respectively.

The symmetry determining equation (\ref{newprolong}) splits with respect to  the $t$-derivatives and $x$-derivatives of $u$ which leads to a overdetermined linear system of determining equations for the infinitesimals $\tau$, $\xi$ and $\eta$ along with the function $f$ satisfying $f'(u)\neq0$. By simplifying this system we obtain that $\tau=\tau(t)$, $\xi=\xi(t,x)$, $\eta=\eta(t,x,u)$ and $f(u)$ must satisfy the following conditions:
\begin{equation}\label{sis1}\begin{array}{r}
u^2 \eta_{xxx}+\eta_x f_u  +\eta_t=0, \\ 3 u \eta_{xu}-3 u \xi_{xx}+4 \eta_x=0,\\

 3 u^2 \eta_{uu}+4 u \eta_{u}-4\eta=0, \\
2 \eta- 3  \xi_x u+ \tau_t u=0, \\

   u^3 \eta_{uuu}+4 u^2 \eta_{uu}+2 u \eta_{u}-2 \eta=0, \\

 3 u^2 \eta_{xuu} +8 u \eta_{xu}+3 \eta_x-4  \xi_{xx}u=0,\\

3 u^3 \eta_{xxu}+4 u^2 \eta_{xx}+u f_{uu} \eta-2 f_u \eta-\xi_{xxx}u^3+2 \xi_x u f_u-\xi_t u =0.

\end{array} \end{equation}

\noindent Maple is used for setting up the determining equation, and the commands ``dsolve'' and ``pdsolve'' to solve the determining system (\ref{sis1}), a complete classification of all solution cases has been achieved. Now, we proceed to show the results.

\begin{theorem}
The point symmetries admitted by the generalized quasilinear KdV equation \eqref{genedp} for arbitrary $f(u)$ are generated by:
\begin{eqnarray}
\label{symm1}
{\bf X}_1&=&\partial_t,\\
\label{symm2}
{\bf X}_2&=&\partial_x.
\end{eqnarray}
Additional point symmetries are admitted by the generalized quasilinear KdV equation \eqref{genedp} in the following cases:
\begin{itemize}
\item[$\bullet$] If $f(u)=f_0 u^n+f_1 u+f_2$, 
\begin{equation}
\label{symm3}
{\bf X}_3=\left(3 n-5 \right)t \partial_t + \left((n-3)x+2 f_1 (n-1)t\right) \partial_x -2 u \partial_u.
\end{equation}
Moreover, if $n=1$, $f_1=0$ besides ${\bf X}_3$ we obtain the following symmetry
\begin{equation}
\label{symm4}
{\bf X}_4=2 \left(x- f_0 t\right) \partial_x+3 u \partial_u.
\end{equation}

\item[$\bullet$] If $f(u)=f_0 u \ln u+f_1 u+f_2$, 
\begin{equation}
\label{symm5}
{\bf X}_5= t \partial_t + \left(x+f_0 t\right) \partial_x+ u \partial_u.
\end{equation}

\item[$\bullet$] If $f(u)=f_0 \ln u+f_1 u+f_2$, 
\begin{equation}
\label{symm6}
{\bf X}_6= 5 t \partial_t+\left(3 x+2 f_1 t\right) \partial_x+ 2 u \partial_u.
\end{equation}

\end{itemize}

In the above symmetries, $n \neq 0$, $f_0\neq 0$, $f_1$ and $f_2$ are arbitrary constants. On the other hand, symmetries (\ref{symm1})-(\ref{symm2}) represent space and time-translations. Symmetries (\ref{symm3})-(\ref{symm6}) are scalings combined with Galilean boosts.
\end{theorem}

\noindent The corresponding symmetry transformation groups are given by
$$
\begin{array}{l}
(t,x,u)_1  \longrightarrow (t+\epsilon,x,u), \\

(t,x,u)_2  \longrightarrow (t,x+\epsilon,u), \\

(t,x,u)_3  \longrightarrow (e^{(3n-5)\epsilon} t,e^{(n-3)\epsilon}(x-f_1 t)+e^{(3n-5)\epsilon}f_1 t,e^{-2\epsilon} u),\\

(t,x,u)_4  \longrightarrow (t,e^{2\epsilon}\left(x-f_0 t\right)+ f_0 t,e^{3\epsilon} u),\\

(t,x,u)_5  \longrightarrow (e^{\epsilon} t,e^{\epsilon}(x+f_0 \epsilon  t),e^{\epsilon} u),\\

(t,x,u)_6  \longrightarrow (e^{5\epsilon} t,e^{3\epsilon}\left(x-f_1 t\right)+ e^{5\epsilon}f_1 t,e^{2\epsilon} u),
\end{array}
$$
\noindent with $\epsilon$ the group parameter.

\section{Conservation laws}
\label{S2}

For the generalized quasilinear KdV equation
\eqref{genedp},
a local conservation law is a space-time divergence expression,
given by
\begin{equation}\label{conslaw}
D_t T+D_x X=0,
\end{equation}
holding on the solution space of equation \eqref{genedp},
where $T$ is the conserved density, and $X$ is the spatial flux,
which are functions of $t$, $x$, $u$ and its derivatives. $D_t$ and $D_x$ denote, respectively, the total derivatives with respect to $t$ and $x$.

If we consider solutions $u(t,x)$ in a given spatial domain $\Omega=(a,b)\subseteq\mathbb{R}$,
then every conservation law yields a conserved integral, given by
\begin{equation}\label{conservedquantity}
\mathcal{C}[u]= \int_a^b T\,dx
\end{equation}
which satisfies the balance equation
\begin{equation}\label{globalconslaw}
\frac{d}{dt}\mathcal{C}[u]= -X \Big|_a^b.
\end{equation}
Similarly, in the case when the spatial domain for solutions is given by $(-\infty,b)$, $(a,\infty)$ or $(-\infty, \infty)$.
The physical meaning of the global equation
\eqref{globalconslaw}
is that
the rate of change of the quantity $\mathcal{C}[u]$
on the spatial domain $\Omega$
is balanced by the net flux through the endpoints.

Since the quasilinear KdV equation \eqref{genedp}
is an evolution equation,
by substituting all $t$ derivatives of $u$ through equation \eqref{genedp},
we can write the conservation law \eqref{conslaw}
in an equivalent form
\begin{equation}\label{conslaw2}
D_t T(t,x,u,u_x,u_{xx},u_{xxx},\dots)+D_x X(t,x,u,u_x,u_{xx},u_{xxx},\dots)=0,
\end{equation}
that will be satisfied on the solution space of  equation \eqref{genedp}.

When considered off of the solution space of equation \eqref{genedp},
the conservation law \eqref{conslaw2}
has an equivalent characteristic form
\begin{equation}\label{charfor}
D_t T+D_x (X-\Psi)
=
\big( u_t+ (u(uu_x)_x+f(u))_x \big)Q
\end{equation}
where
\begin{equation}
\Psi = \big( u_t+ (u(uu_x)_x+f(u))_x \big)\hat{E}_{u_x}(T)
+D_x\big( u_t+ (u(uu_x)_x+f(u))_x\big)\hat{E}_{u_{xx}}(T)
+\cdots
\end{equation}
vanishes on the solution space of equation \eqref{genedp},
and where $Q$ is a multiplier given by
\begin{equation}\label{multQ}
Q=\hat{E}_u(T),
\end{equation}
where $\hat{E}_u$ is the spatial Euler operator with respect to the variable $u$ \cite{ancorev,blumanchevianco,PJO}.
To find the conservation laws of equation \eqref{genedp}
we need to find the multipliers that solve the characteristic equation \eqref{charfor}.
Since the right hand side of equation \eqref{charfor}
must be a divergence expression,
all multipliers can be found by solving
the determining equation holding off of solutions, given by
\begin{equation}\label{deteqn}
E_u\big((u_t+ (u(uu_x)_x+f(u))_x) Q\big)=0.
\end{equation}

Here $E_u$ is the Euler operator with respect to the variable $u$,
which is characterized by annihilating total derivatives expressions
\cite{ancorev,blumanchevianco,PJO}.
The determining equation \eqref{deteqn}
can be solved by splitting with respect to the variables that do not appear in $Q$,
yielding an overdetermined linear system of equations for $Q$ and the arbitrary function $f(u)$.

The most interesting conservation laws
for dispersive wave equations
that are interesting
for physics
usually
correspond to low-order multipliers,
meanwhile
high-order multipliers are related to
the integrability of the equation \cite{Anc16,ancorev}.
Therefore
the dispersive quasilinear KdV equation \eqref{genedp}
has low-order multipliers $Q$ given by
\begin{equation}\label{multr}
Q(t,x,u,u_x,u_{xx}).
\end{equation}
The dependence of $Q$ can be found by considering the variables $u$ that can be differentiated to obtain a leading derivative of equation \eqref{genedp}.
Thus,
in order to obtain a general classification of
low-order multipliers \eqref{multr} for equation \eqref{genedp}
we solve the determining equation \eqref{deteqn}.
We carry out this classification by using the Maple package ``rifsimp'' and ``pdsolve''.

The solutions with $f'(u)\neq 0$ are given by:
\begin{itemize}
\item $f(u)$ arbitrary function,
\begin{align}
& Q_1 = 1,
\label{mult1}
\\
& Q_2 = u,\\
& Q_3 = -u^2 u_{xx} - u u_x^2 - f(u).
\end{align}
\item $f(u)=f_0 u^5 + f_1 u + f_2$, where $f_0$, $f_1$, $f_2$ are arbitrary constants with $f_0 f_1\neq 0$.
$$
Q_1, \quad
Q_2, \quad
Q_3,
$$
\begin{equation}\label{mult4}
Q_4 = 2t(u^2 u_{xx} + u u_x^2 + f_0 u^5) + \tfrac{2}{5}(f_1 t-x)u.
\end{equation}
\end{itemize}

Next, for each of the multipliers
\eqref{mult1}--\eqref{mult4}
we determine the corresponding conserved density $T$ and flux $X$.
This is straightforwardly done by integrating the characteristic equation \eqref{charfor} or by using a homotopy integral formula \cite{ancorev}.
We summarize the results in the following theorem.

\begin{theorem}
(i) The low-order local conservation laws of
the generalized quasilinear KdV equation \eqref{genedp}
for arbitrary $f'(u)\neq 0$ are given by:
\begin{equation}\label{TX1}
\begin{split}
T_1 =& u,\\
X_1 =& u^2 u_{xx} + u u_x^2 + f(u).
\end{split}
\end{equation}
\begin{equation}\label{TX2}
\begin{split}
T_2 =& \tfrac{1}{2} u^2,\\
X_2 =& u^3 u_{xx} + \tfrac{1}{2} u^2 u_x^2 + \textstyle\int u f'(u)\,du.
\end{split}
\end{equation}
\begin{equation}\label{TX3}
\begin{split}
T_3 =&  \tfrac{1}{2} u^2 u_x^2 - \textstyle\int f(u)\,du,\\
X_3 =& -\tfrac{1}{2} u^4 u_{xx}^2 - u^3 u_x^2 u_{xx}
- \tfrac{1}{2} u^2 u_x^4  - u^2 u_t u_x - f(u) u^2 u_{xx}
\\
& \qquad  - f(u)u u_x^2 - \tfrac{1}{2} f(u)^2.
\end{split}
\end{equation}
(ii) The generalized quasilinear KdV equation
\eqref{genedp} only admits an additional low-order local conservation law
for $f(u)=f_0 u^5 + f_1 u + f_2$, where $f_0$, $f_1$, $f_2$ are arbitrary constants with $f_0 f_1\neq 0$, given by:
\begin{equation}\label{TX4}
\begin{split}
T_4 =& \tfrac{1}{3} f_0 t u^6  -t u^2 u_x^2
+ \tfrac{1}{5} (f_1 t-x) u^2,\\
X_4 =& 2 f_0 t u^7 u_{xx} + t u^4 u_{xx}^2
+ 2t u^3 u_x^2 u_{xx} + \tfrac{2}{5} f_1 t u^3 u_{xx}
- \tfrac{2}{5} x u^3 u_{xx} + \tfrac{2}{5} u^3 u_x
\\
& \qquad
+ t u^2 u_x^4 - \tfrac{1}{5} x u^2 u_x^2
+ 2t u^2 u_t u_x + 2 f_0 t u^6 u_x^2
+ \tfrac{6}{5} f_1 t u^2 u_x^2 + f_0^2 t u^{10}
\\
& \qquad\qquad
+ \tfrac{2}{3} f_0 f_1 t u^6 - \tfrac{1}{3} f_0 x u^6
+ \tfrac{1}{5} f_1^2 t u^2
- \tfrac{1}{5} f_1 x u^2.
\end{split}
\end{equation}
\end{theorem}

Now we give the physical meaning of the conservation laws for solutions $u(t,x)$ of the generalized quasilinear KdV equation \eqref{genedp}.
Conservation law \eqref{TX1} is equation \eqref{genedp} itself,
describing conservation of mass
\begin{equation}
\mathcal{M}[u] = \int_{\Omega} u \, dx.
\end{equation}
Conservation law \eqref{TX2} describes
conservation of momentum
\begin{equation}
\mathcal{P}[u] = \int_{\Omega}\tfrac{1}{2} u^2 \, dx.
\end{equation}
Conservation law \eqref{TX3} describes
conservation of energy
\begin{equation}
\mathcal{E}[u] = \int_{\Omega} \Big( \tfrac{1}{2} u^2 u_x^2 - \textstyle\int f(u)\,du\Big) \, dx.
\end{equation}
Finally, conservation law \eqref{TX4} yields
the conserved quantity
\begin{equation}
\mathcal{J}[u] = \int_{\Omega} \Big( \tfrac{1}{3} f_0 t u^6  -t u^2 u_x^2 + \tfrac{1}{5} (f_1 t-x) u^2\Big) \, dx,
\end{equation}
for solutions $u(t,x)$ of the quasilinear KdV equation \eqref{genedp} with $f(u)=f_0 u^5 + f_1 u + f_2$.

 \section{Travelling wave reductions}\label{ps}
For the generalized quasilinear KdV equation \eqref{genedp} we obtain travelling wave reductions through the change of variables
\begin{equation}\label{trans}\begin{array}{lll} z=\mu x-\lambda t,&\hspace{0.5cm}
&u=w(z),\end{array}\end{equation} where $w(z)$ satisfies
\begin{equation}\label{edor}
\mu^3\,w^2\,w'''+4\,\mu^3\,w\,w'\,w''-\lambda \,w'+\mu^3\,\left(w'\right)^3+f_{w}\,\mu\,w'=0.
\end{equation}
By integrating equation \eqref{edor} with respect to $z$ we get
\begin{equation}\label{edo0}
\mu^3\,w^2\,w''-\lambda w+\mu^3\,w\,\left(w'\right)^2+ \mu f+k=0,
\end{equation} where $k$ is a constant of integration.

\subsection{Integrating factors and first integrals}

Next we focus our attention on obtaining integrating factors of ODE \eqref{edo0}. From \cite{ AB98, BA} for any $n$th-order ODE finding an integrating factor is equivalent to finding a first integral. Furthermore, a first integral of equation (\ref{edo0}) yields a quadrature which reduces (\ref{edo0}) to a 1st-order ODE in terms of the original variables $z$, $w$, $w'$.

An integrating factor of equation (\ref{edo0}) is a function  $Q(z,w,w')\neq 0$ which verifies the characteristic equation \cite{BA}
\begin{eqnarray}\label{chareqn}
D_z \tilde X&=& \left(\mu^3\,w^2\,w''-\lambda w+\mu^3\,w\,\left(w'\right)^2+ \mu f(w)+k\right)Q, 
\end{eqnarray} where the spatial flux $\tilde X$ is  a function of $z$, $w$, and $z$-derivatives of $w$.

Moreover, all first integrals of ODE (\ref{edo0}) derive from integrating factors verifying the characteristic equation (\ref{chareqn}) for any function $w(z)$. The determining equation to obtain all integrating factors is
\begin{eqnarray}\label{multreqn}
E_w\Big( (\mu^3\,w^2\,w''-\lambda w+\mu^3\,w\,\left(w'\right)^2+ \mu f+k)Q \Big) &=&0,
\end{eqnarray}
where $E_w$ represents the variational derivative with respect to $w$. This equation must hold off of the set of solutions of equation (\ref{edo0}). Once the integrating factors are found, the corresponding non-trivial first integrals are obtained by using a line integral \cite{BA}.

\noindent The determining equation  (\ref{multreqn}) splits with respect to the variable $w''$ and its derivatives. This yields a linear determining system for $Q$ from which we obtain the integrating factor $Q=w'$. For any integrating factor $Q(z,w,w')$ the first integral formula yields \cite{AB98}

\begin{eqnarray}\label{sol}
 \tfrac{1}{2}{w'}^{2}{\mu}^{3}{w}^{2}+\int \!(\mu f \left( w \right) -\lambda w+k)\,{\rm d}w=c,\quad \mbox{with $c$ constant}.
\end{eqnarray}

\subsection{Some exact solutions}

Equation \eqref{sol} can be written in the form
\begin{equation}\label{sols}
w'^2=-{\frac{2}{\mu^2\,w^2}}\,\int {f\left(w\right)}{\;dw}-{\frac{2\,k}{
 \mu^3\,w}}+{\frac{2\,c}{\mu^3\,w^2}}+{\frac{ \lambda}{\mu^3}}.
\end{equation}

\noindent Equation \eqref{sols} with
$$f(w)=-3 {  C_4} \mu^2 w^5-{ \frac{5 {  C_3} \mu^2 w^4}{2}}-2 {  C_2} \mu^2 w^3-{\frac{3 {  C_1} \mu^2 w^2}{2}}-{  C_0}  \mu^2 w+{\frac{{  \lambda} w}{\mu}}-{\frac{k}{\mu}},$$
\noindent is the  equation for the Jacobi elliptic function

\begin{equation}\label{jac}
 w'^2=C_4w^4+C_3w^3+C_2w^2+C_1w+C_0,
\end{equation}
where $C_i$, $i=0\ldots,4$ are constants.

\noindent Equation \eqref{sols} with
$$f(w)=-10 \mu^2 w^4+{\frac{3 {  g_2} \mu^2 w^2}{2}}+{  g_3} \mu
 ^2 w+{{{  \lambda} w}\over{\mu}}-{{k}\over{\mu}},$$  is the equation for the Weierstrass elliptic function
 \begin{equation}\label{wei}
 w'^2=4w^3-g_2w-g_3.
\end{equation}

If $C_0=C_1=C_4=0$, $C_3=-4$ and $C_2=4$ then equation \eqref{jac} admits the solution $w={\rm sech}^2(z)$. Consequently, for
$$f(w)=10\,\mu^2\,w^4-8\,\mu^2\,w^3+{{{ \lambda}\,w}\over{\mu}}-{{k }\over{\mu}},$$
equation \eqref{genedp} admits the solution
$$u(x,t)={\rm sech}^2(\mu x-\lambda t).$$
This solution represents a soliton solution with speed $\frac{\lambda}{\mu}$ and moving coordinate $\mu x - \lambda t$.
\begin{figure}[htp]
\begin{center}
  \includegraphics[width=3in]{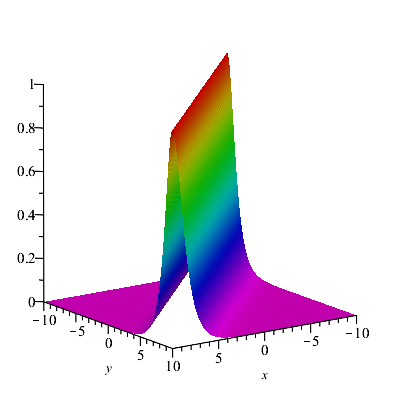}\\
  \caption{Solution $u(x,t)={\rm sech}^2(x-t)$ of Eq. (\ref{genedp})}\label{sol1}
\end{center}
\end{figure}

Non-singular solutions are interesting in physical sciences. From \cite{Yom} we provide some sets of these types of solutions of equation \eqref{jac} considering $C_1=C_3=0$:
\begin{enumerate}
\item If $C_0=a^2$, $C_2=2m^2-1$ and $C_4={{m^4-m^2}\over{a^2}}$
$$w_1(z)={{a\,{\rm  sn}\left(z , m\right)}\over{\sqrt{1-m^2\,
 {\rm  sn}^2\left(z , m\right)}}}.$$

\item If $C_0=a^2$, $C_2=-(m^2+1)$ and $C_4=\frac{m^2}{a^2}$ and $m^2<1$
$$w_2(z)={{a\,{\rm sn}\left(z , m\right)}\over{\sqrt{-{{m^2\,
 {\rm cn}^2\left(z , m\right)-m^2+1}\over{m^2-1}}}}}.$$

\item If $C_0=-a^2(m^2-1)$, $C_2=-{{3\,m^6-3\,m^4-m^2-1}\over{3\,m^4-1}}$ and $C_4=-{{m^2\,\left(8\,m^4+m^2-1\right)}\over{a^2\,\left(m^2+1\right)^2}}$
$$w_3(z)= {{a\,{\rm cn}\left(z , m\right)}\over{\sqrt{d\,
 {\rm cn}^2\left(z , m\right)+1}}}.$$

\item If $C_0=a^2$, $C_2=-(m^2+1)$ and $C_4=\tfrac{m^2}{a^2}$
$$w_4(z)={{a\,{\rm cn}\left(z , m\right)}\over{\sqrt{1-m^2\,
 {\rm sn}^2\left(z , m\right)}}}.$$

\item If $C_0={{a^2\,\left(m-1\right)\,\left(m+1\right)}\over{m^2}}$, $C_2=-{{m^2+1}\over{m^2}}$ and $C_4={{1}\over{a^2\,\left(m-1\right)\,\left(m+1\right)}}$
    $$w_5(z)={{a\,{\rm dn}\left(z , m\right)}\over{\sqrt{{{
 {\rm dn}^2\left(z , m\right)+m^2-1}\over{m^2-1}}}}}.$$
 \end{enumerate}

In the previous solutions, ${\rm sn}\left(z , m\right)$ is the Jacobi elliptic sine function, ${\rm cn}\left(z , m\right)$ is the Jacobi elliptic cosine function, and ${\rm dn}\left(z , m\right)$ is the Jacobi elliptic function of the third
kind. Taking into account \eqref{trans} we derive exact solutions for equation \eqref{genedp}.

For example, for $C_0=C_2=1$, $C_1=C_3=C_4=0$, and by using both ${\rm sn}\left(z , 1\right)={\rm tanh}(z)$ and  ${\rm cn}\left(z ,1\right)={\rm sech}(z)$ along with the relation ${\rm sn}^2(z)+{\rm cn}^2(z)=1$, equation \eqref{jac} admits the solution $w_1={\rm sinh}(z)$. Therefore, for
$$f(w)=-2 \mu^2 w^3-  \mu^2 w+{\frac{{  \lambda} w}{\mu}}-{\frac{k}{\mu}},$$
equation \eqref{genedp} admits the solution
$$u(x,t)={\rm sinh}( \mu x-\lambda t).$$

\begin{figure}[htp]
\begin{center}
  \includegraphics[width=3in]{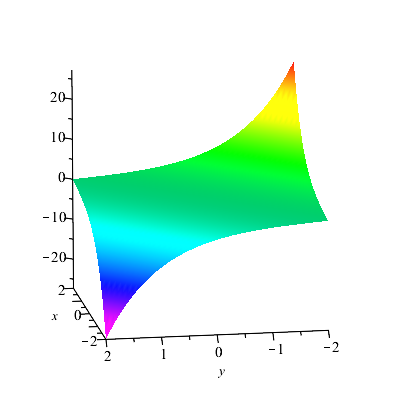}\\
  \caption{Solution $u(x,t)={\rm sinh}(x-t)$ of Eq. \eqref{genedp}}\label{sol2}
  \end{center}
\end{figure}

On the other hand, for $C_0=1$, $C_2=-1$, $C_1=C_3=C_4=0$, and since ${\rm sn}\left(z , 0\right)={\rm sin}(z)$, equation \eqref{jac} admits the solution $w_2={\rm sin}(z)$. Therefore, for
$$f(w)=2 \mu^2 w^3-  \mu^2 w+{\frac{{  \lambda} w}{\mu}}-{\frac{k}{\mu}}$$
equation \eqref{genedp} admits the periodic solution
$$u(x,t)={\rm sin}( \mu x-\lambda t).$$

\begin{figure}[htp]
\begin{center}
  \includegraphics[width=3in]{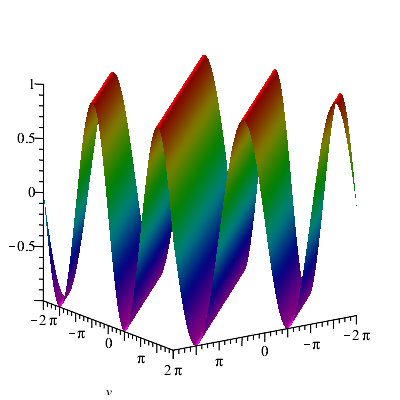}\\
  \caption{Solution $u(x,t)={\rm sin}(x-t)$ of Eq. \eqref{genedp}}\label{sol3}
  \end{center}
\end{figure}

\section{Conclusions}
\label{S3}

In this work, we have analysed a generalized quasilinear KdV equation with degenerate dispersion containing an arbitrary function of the dependent variable from the point of view of symmetries and conservation laws. To begin with, we have derived a complete classification of all point symmetries that equation (\ref{genedp}) admits as well as provided a physical interpretation of them.
Next, we have obtained all low-order local conservation laws of the quasilinear KdV equation \eqref{genedp}. In addition, we have given the physical description of the conserved quantities.
Finally, some exact solutions have been derived through the travelling wave reduction and by the quadrature obtained by using first integrals.

\section*{Acknowledgments}
The authors acknowledge the financial support from Junta de Andaluc\'{i}a group FQM-201 and they express their sincere gratitude to the Plan Propio de Investigaci\'{o}n de la Universidad de C\'{a}diz.

\providecommand{\href}[2]{#2}
\providecommand{\arxiv}[1]{\href{http://arxiv.org/abs/#1}{arXiv:#1}}
\providecommand{\url}[1]{\texttt{#1}}
\providecommand{\urlprefix}{URL }


\begin{thebibliography}{10}


\bibitem{Alo}
\newblock  L.~M. Alonso,
\newblock On the Noether map,
\newblock \emph{Letters in Mathematical Physics},  {\bf 3} (1979), 419--424.

\bibitem{Amb}
\newblock  D. Ambrose and J. Wright,
\newblock Traveling waves and weak solutions for an equation with degenerate dispersion,
\newblock \emph{Proceedings of the American Mathematical Society},  {\bf 141} (2013), 3825--3838.

\bibitem{Anc16}
\newblock S.~C. Anco,
\newblock Symmetry properties of conservation laws,
\newblock \emph{International Journal of Modern Physics B}, \textbf{30} (2016).

\bibitem{ancorev}
\newblock S.~C. Anco,
\newblock Generalization of Noether's Theorem in Modern Form to Non-variational Partial Differential Equations,
in \emph{Recent Progress and Modern Challenges in Applied Mathematics, Modeling and Computational Science} \textbf{79},  Springer, (2017), 119--182.

\bibitem{anco0}
\newblock S.~C. Anco and G.~W. Bluman,
\newblock Direct construction of conservation laws from field equations,
\newblock \emph{Physical Review Letters}, \textbf{78} (1997), 2869--2873.

\bibitem{AB98}
\newblock S.~C. Anco and G.~W. Bluman,
\newblock Integrating factors and first integrals for ordinary differential equations,
\newblock \emph{European Journal of Applied Mathematics}, \textbf{9} (1998), 245--259.

\bibitem{anco1}
\newblock S.~C. Anco and G.~W. Bluman,
\newblock Direct constrution method for conservation laws of partial differential equations part 1: Examples of conservation law classifications,
\newblock \emph{European Journal of Applied Mathematics}, \textbf{5} (2002), 545--566.


\bibitem{anco}
\newblock S.~C. Anco and G.~W. Bluman,
\newblock Direct construction method for conservation laws of partial differential equations part 2: General treatment,
\newblock \emph{European Journal of Applied Mathematics.}, \textbf{5} (2002), 567--585.

\bibitem{BA}
\newblock G.~ W.~Bluman, S.~Anco,
\newblock \emph{Symmetry and Integration Methods for Differential Equations},
\newblock Applied Mathematical Sciences series \textbf{154}, Springer, 2002.

\bibitem{blumanchevianco}
\newblock G.~W. Bluman, A.~F. Cheviakov and S.C.~Anco,
\newblock \emph{Applications of symmetry methods to partial differential equations},
\newblock Springer-New York, 2010.

\bibitem{bluco}
\newblock G.~W. Bluman and J.~Cole,
\newblock General similarity solution of the heat equation,
\newblock \emph{Journal of Mathematics and Mechanics}, \textbf{18} (1969), 1025--1042.

\bibitem{bluku}
\newblock G.~W. Bluman and S.~Kumei,
\newblock On the remarkable nonlinear diffusion equation,
\newblock \emph{Journal of Mathematical Physics}, \textbf{21} (1980), 1019--1023.

\bibitem{blu2}
\newblock G.~W. Bluman, S.~Kumei and G.~J. Reid,
\newblock New classes of symmetries for partial differential equations,
\newblock \emph{Journal of Mathematics and Mechanics}, \textbf{29} (1988), 806--811.

\bibitem{bru1}
\newblock M.~S. Bruz\'{o}n and T.~M. Garrido,
\newblock Symmetries and conservation laws of a KdV6 equation,
\newblock \emph{Discrete and Continuous Dynamical Systems Series S}, \textbf{11} (2018), 631--641.

\bibitem{bru2}
\newblock M.~S. Bruz\'{o}n, E. Recio, T.~M. Garrido, A.~P. M\'{a}rquez and R. de la Rosa,
\newblock On the similarity solutions and conservation laws of the Cooper-Shepard-Sodano equation,
\newblock \emph{Mathematical Methods in the Applied Sciences}, \textbf{41} (2018), 7325--7332.

\bibitem{bru3}
\newblock M.~S. Bruz\'{o}n, M.~L. Gandarias and R. de la Rosa,
\newblock An Overview of the Generalized Gardner Equation: Symmetry Groups and Conservation Laws, in
\newblock \emph{A Mathematical Modeling Approach from Nonlinear Dynamics to Complex Systems}, (2019), 7--27.

\bibitem{coop}
\newblock F. Cooper, H. Shepard and P. Sodano,
\newblock Solitary waves in a class of generalized Korteweg-de Vries equations,
\newblock \emph{Physical Review E}, \textbf{48} (1993), 4027--4032.

\bibitem{ger}
\newblock P. Germain, B. Harrop-Griffiths and J. Marzuola,
\newblock Existence and uniqueness of solutions for a quasilinear KdV equation with degenerate dispersion, preprint,
\newblock arXiv: 1801.00420v1.

\bibitem{KK}
\newblock A.~Karasu-Kalkanli, A.~Karasu, A.~Sakovich, S.~Sakovich and  R.~Turhan,
\newblock A new integrable generalization of the Korteweg de Vries equation,
\newblock \emph{Journal of Mathematical Physics}, \textbf{49} (2008).

\bibitem{kudry1}
\newblock  N.~Kudryashov,
\newblock  On `new travelling wave solutions' of the KdV and the KdV Burgers equations.
\newblock \emph{Communications in Nonlinear Science and Numerical Simulation}, \textbf{14}  (2009), 1891--1900.

\bibitem{PJO}
\newblock P.~J. Olver,
\newblock \emph{Applications of Lie groups to differential equations},
\newblock Springer-Verlag,  1986.

\bibitem{recio}
\newblock E. Recio and S.~C. Anco,
\newblock Conservation laws and symmetries of radial generalized nonlinear p-Laplacian evolution equations,
\newblock \emph{Journal of Mathematical Analysis and Applications}, \textbf{452} (2017), 1229--1261.

\bibitem{ros}
\newblock P. Rosenau and J.~M. Hyman,
\newblock Compactons: solitons with finite wavelength,
\newblock \emph{Physical Review Letters}, \textbf{70} (1993), 564--567.

\bibitem{sat}
\newblock J.~Satsuma and R.~Hirota,
\newblock A coupled KdV equation is one case of the four--reduction of the KP hierarchy,
\newblock \emph{Journal of the Physical Society of Japan}, \textbf{51} (1982), 3390--3397.

\bibitem{saw}
\newblock K.~Sawada and T.~Kotera,
\newblock A method for finding N-soliton solutions of the KdV equation and KdV-like equation,
\newblock \emph{Progress of Theoretical Physics}, \textbf{51} (1974), 1355--1367.

\bibitem{WTC}
\newblock J.~Weiss, M.~Tabor and G.~Carnevale,
\newblock The painlev\`e property for partial differential equations,
\newblock \emph{Journal of Mathematical Physics}, \textbf{24} (1983), 522--526.

\bibitem{Yom}
\newblock G.~A.~Zakeri, E.~Yomba,
\newblock  Exact solutions of a generalized autonomous Duffing-type equation,
\newblock \emph{Applied Mathematical Modelling}, \textbf{39} (2015), 4607--4616.

\end{thebibliography}
\end{document}